%% file: AMEE.tex
\documentclass[aps,prl,twocolumn,showpacs,longbibliography,superscriptaddress]{revtex4-2}
\usepackage{tikz} 
\usetikzlibrary{shapes,arrows,positioning,automata,backgrounds,calc,er,patterns}
\usepackage{graphicx}
\usepackage{dcolumn}
\usepackage{bm}
\usepackage{amsmath}
\usepackage{amssymb}
\usepackage{amsthm}
\usepackage{amsfonts}
\usepackage{enumerate}
\usepackage{siunitx}
\usepackage[colorlinks=true,urlcolor=blue,citecolor=blue,linkcolor=blue]{hyperref}

\definecolor{purple}{rgb}{0.8,0,0.6}
\definecolor{darkgreen}{rgb}{0.00,0.6,0.00}

\begin{document}

\title{Axial Magnetoelectric Effect in Dirac semimetals
}

\newcommand{\affiliationNordita}{Nordita, KTH Royal Institute of Technology and Stockholm University, Roslagstullsbacken 23, SE-106 91 Stockholm, Sweden}
\newcommand{\affiliationConnecticut}{Department of Physics and Institute for Materials Science, University of Connecticut, Storrs, Connecticut 06269, USA}
\newcommand{\affiliationYale}{Department of Physics, Yale University, New Haven, CT 06520, USA}

\author{Long Liang}
\affiliation{\affiliationNordita}
\author{P. O. Sukhachov}
\affiliation{\affiliationYale}
\author{A. V. Balatsky}
\affiliation{\affiliationNordita}
\affiliation{\affiliationConnecticut}

\begin{abstract}
We propose a mechanism to generate a static magnetization via {\em axial magnetoelectric effect} (AMEE). Magnetization ${\bf M} \sim {\bf E}_5(\omega) \times {\bf E}_5^{*}(\omega)$ appears as a result of the transfer of the angular momentum of the axial electric field ${\bf E}_5(t)$ into the magnetic moment in Dirac and Weyl semimetals.
We point out similarities and differences between the proposed AMEE and a conventional inverse Faraday effect (IFE). 
As an example, we estimated the AMEE generated by circularly polarized acoustic waves and find it to be on the scale of microgauss for gigahertz frequency sound. In contrast to a conventional IFE, magnetization rises linearly at small frequencies and fixed sound intensity as well as demonstrates a nonmonotonic peak behavior for the AMEE.
The effect provides a way to investigate unusual axial electromagnetic fields via conventional magnetometry techniques.
\end{abstract}
\maketitle

\emph{Introduction.--}
We now witness the surge of interest in quantum  matter with unusual relativisticlike dispersion relation~\cite{Franz:book-2013,Vafek-Vishwanath:rev-2014,Wehling-Balatsky:rev-2014,Armitage:rev-2018}. 
We mention bosonic (acoustic waves, photonic crystals, and magnons) and fermionic ($A$-phase of superfluid helium-3, $d$-wave superconductors, graphene, topological insulators)  Dirac and Weyl  materials. The key topological  features for all of them are the linear energy spectrum and the presence of several gapless nodes (valleys) that enable exotic artificial fields, unique for Dirac materials. 

Dynamics of excitations in Dirac and Weyl semimetals and their interplay with electromagnetic fields is a rapidly developing research area. 
We focus on the inverse Faraday effect (IFE)~\cite{Pitaevskii:1960,Pershan:1963,Ziel-Malmstrom:1965,Pershan-Malmstrom:1966,PhysRevB.89.014413} that is a well-known example of light-matter interaction. 
In this case, the angular momentum from a dynamical electromagnetic (photon) field $\mathbf{E}(t)$ is transferred into the magnetic moment of electrons, leading to a static magnetization $\mathbf{M} \sim \mathbf{E}(\omega)\times \mathbf{E}^*(\omega)$. 
This effect was recently investigated for Dirac and Weyl semimetals in Refs.~\cite{Taguchi-Tanaka:2016,Zyuzin-Zyuzin:2018,PhysRevB.101.174429,kawaguchi2020giant,Gao-Xiao:IFE-2020}.

In this Letter, we propose a  mechanism to generate a static magnetization in Dirac and Weyl semimetals through dynamical axial gauge fields, whereby the angular momentum is transferred from the axial or pseudo-electric field $\mathbf{E}_5(t)$ into the magnetic moment of electrons 
\begin{equation}
\mathbf{M} \sim \mathbf{E}_5(\omega)\times\mathbf{E}_5^*(\omega).
\end{equation}
Therefore, we call this phenomenon the {\em axial magnetoelectric} effect (AMEE). 
The AMEE is qualitatively different from the IFE. 
Synthetic nature of the pseudoelectric field in Dirac and Weyl semimetals means that one can induce magnetization using phonons without any electromagnetic fields. This situation reminds the difference between the Aharonov-Bohm (AB)~\cite{AB_effect} and Aharonov-Casher (AC)~\cite{AC_effect} effects, where the AB and AC gauge potentials, while similar, have a different origin.

The origin of the pseudo-electromagnetic fields is rooted in the relativisticlike energy spectrum and  topology of Dirac matter. In the case of Dirac semimetals, axial gauge fields could be generated via strains~\cite{Suzuura-Ando:2002,Vozmediano-Guinea:2010,Zhou_2013,PhysRevLett.115.177202,Pikulin-Franz:2016,Grushin:2016,Ilan-Pikulin:rev-2019} as well as magnetization textures~\cite{Liu-Qiu:2012,Araki:rev-2020}. In graphene, pseudomagnetic fields were observed experimentally in Refs.~\cite{Levy-Crommie:2010,Liu-Loh:2018,Nigge-Damascelli:2019,Hsu-Yeh:2020}.
Previous work focuses mostly on static axial fields. The studies related to dynamics of these fields quickly become an active research area. In particular, we mention ultrasonic attenuation~\cite{Spivak:2016,Pikulin-Franz:2016,Rinkel-Garate:2019}, acoustogalvanic effect~\cite{PhysRevLett.124.126602}, and torsion-induced chiral magnetic current~\cite{Gao-Kharzeev:2020}. 
The circularly polarized sound can be generated by acoustic transducers coupled to acoustic quarter-waveplates~\cite{Lyamov:book}, see also Refs.~\cite{Auld:1966,Vaart:1966,Lemanov:1971}.
In addition to the acoustic method, we also point out recent work on generation of the circularly-polarized phonons~\cite{Juraschek-Spaldin:2019}.
It was shown that circularly-polarized phonons can be induced
optically~\footnote{In the case where circularly-polarized phonons are induced optically, a conventional IFE should be also taken into account. Since the decay depths and rates of photons and phonons are different, the AMEE and the IFE can be distinguished in nonlocal and time-dependent probes.} in Cd$_3$As$_2$~\cite{Cheng-Armitage-Cd3As2:2020}. 
Therefore, the proposed AMEE is timely and instrumental in the elucidation of the  axial fields in three-dimensional semimetals.

\begin{figure}[t]
	\includegraphics[width=0.4\textwidth]{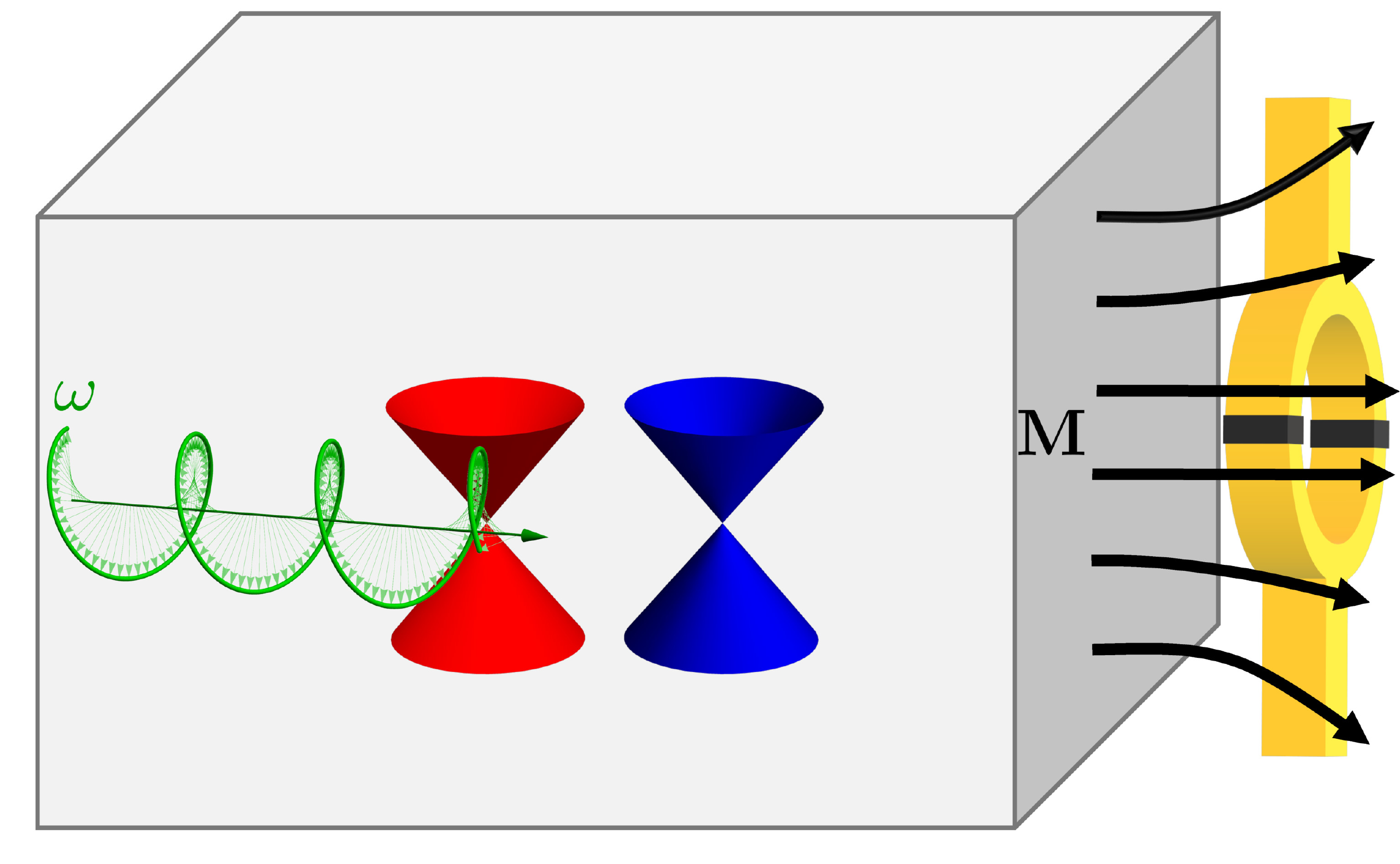}
	\caption{Schematic model setup for the axial magnetoelectric effect. Circularly polarized axial electric fields lead to the static  magnetization $\mathbf{M} \sim \mathbf{E}_5(\omega)\times \mathbf{E}_5^*(\omega)$ of Dirac semimetals. The resulting magnetization is picked up by a magnetometer such as SQUID.}
	\label{Fig:model}
\end{figure}

The AMEE is universal in the sense of measuring the response of nodal states to dynamic axial fields regardless of their origin. Therefore, it could be realized in various experimental setups where the position of the Weyl nodes can be manipulated (for a recent study of the manipulation of the Weyl nodes via light pulses, see Ref.~\cite{Sie-Lindenberg:2019}). 

By using a dynamical deformation as a characteristic example, we found that the induced magnetization scales as $M\propto \mu \omega$ for small frequencies $\omega$ at fixed sound intensity. This is in drastic contrast to the conventional IFE, where the magnetization scales as $M\propto 1/\omega$ in the dirty limit. The explanation of this scaling is simple and follows from the nature of the axial electric field. As in case of conventional IFE, the induced magnetization scales as the second power of the effective axial electric field ${\bf E}_5$.
Because ${\bf E}_5$ is proportional to the time derivative of the gauge potentials, which in turn are determined by the strain tensor $u_{ij}\sim\omega$, this results in two powers of frequency difference when the sound intensity $I\sim\omega^2$ is fixed. Hence, we expect that AMEE has the  
overall scaling $\sim \omega^2$ relative to the IFE for conventional metals. Furthermore, our estimates show that the magnetization induced via the AMEE is within the reach of the modern magnetometry techniques such as SQUID. The corresponding model setup is schematically shown in Fig.~\ref{Fig:model}.

\begin{figure}
	\includegraphics[width=0.2\textwidth]{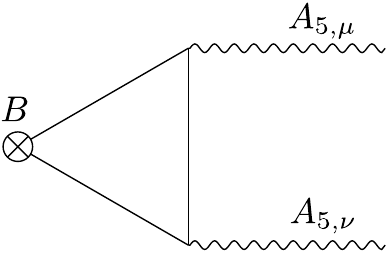}
	\caption{The Feynman diagram used to calculate the AMEE.
	The cross denotes an external magnetic field. Wavy lines represent the axial gauge fields.}\label{Fig:FeynmanDiagram}
\end{figure}

\emph{General theory.--} For weak deviations from equilibrium, the magnetization is obtained as a functional derivative of the effective action $S$ with respect to a magnetic field $\mathbf{M}=-\lim_{B\to0}\delta S/\delta\mathbf{B}$. See Fig.~\ref{Fig:FeynmanDiagram} for the corresponding Feynman diagram. 
Here $A_{5,\mu}=\left\{A_{5,0},-\mathbf{A}_5\right\}$ is the axial gauge field which gives rise to a time-dependent axial electric field $\mathbf{E}_5=-\partial_t \mathbf{A}_5-\bm{\nabla}A_{5,0}$. 
While the diagram in Fig.~\ref{Fig:FeynmanDiagram} resembles that for the chiral anomaly, it has a different structure, where the external lines correspond to two axial and one vector gauge fields. Therefore, the chiral anomaly action does not contribute to the magnetization. This allows us to calculate the effect separately for each Weyl node. We further assume that the wave vector of the chiral electric field is small enough to ignore the internode transitions.

In general, there could be first-order effects, i.e., $M\sim E_5$. They, however, produce an oscillating magnetization that averages to zero for dynamical axial electric fields as well as requires an explicit breakdown of the time-reversal symmetry (TRS). 
Therefore, we focus on a second-order rectification effect, which gives a static magnetization even in a time-reversal symmetric system.
This fact makes the AMEE universal and attractive from a practical point of view. For example, while the first-order effects cancel in Dirac semimetals, the AMEE survives.

Without the loss of generality, we assume that $\mathbf{E}_5$ is polarized in the $x-y$ plane. Then, the magnetization for a single node is
\begin{equation}
\label{M-def}
M_z=-i\chi(\omega,\mathbf{q})(\mathbf{E}_{5,\omega,\mathbf{q}}\times\mathbf{E}_{5,-\omega,-\mathbf{q}})_z 
\end{equation}
with the response function $\chi(\omega,\mathbf{q})$  given by
\begin{widetext}
\begin{eqnarray}
\chi(\omega,\mathbf{q})&=&\frac{e^3}{4\omega^2}\Bigg\{\sum_m \int\mathrm{d}\mathbf{k}~I_{mmm}(\omega,\mathbf{q})
\left(
v^m_y v^m_x\partial_{k_x}v^m_y-v^m_yv^m_y \partial_{k_x}v^m_x
\right)\nonumber\\
&&+\sum_{l\ne m} \int\mathrm{d}\mathbf{k}~I_{mml}(\omega,\mathbf{q})(\epsilon_l-\epsilon_m)\left[(\epsilon_l-\epsilon_m)^2(\Omega^{m}_{xy})^2+ v^m_xv^m_xg^m_{yy}-v^m_xv^m_yg^m_{xy} \right]+(x\leftrightarrow y)\Bigg\}.\label{Eq:IFE}
\end{eqnarray}
\end{widetext}
Here the expression for $I_{mnl}(\omega,\mathbf{q})$ is given in the Supplemental Material (SM)~\cite{SI}, $m=\pm$ denotes the electron and hole bands, $\epsilon_{\pm}=\pm\sqrt{v^2_ik^2_i}$ is the energy dispersion, $v^m_{i}=\partial_{k_i}\epsilon_{m}$ is the quasiparticle velocity, $\Omega^m_{xy}$ and $g^m_{ij}$ are the Berry curvature~\cite{Berry:1984} and the quantum metric~\cite{Provost1980} of the band, respectively (see SM~\cite{SI} for the explicit expressions). The first term in the curly brackets is the intraband contribution, which depends only on the  dispersion. The second term originates from interband processes. In this Letter, we focus on the orbital contribution to the magnetization. The spin contribution is estimated to be weak for intraband processes, $\omega\ll2|\mu|$ (see also Ref.~\cite{Gao-Xiao:IFE-2020} for the IFE results).

A minimal Weyl Hamiltonian
\begin{eqnarray}
H=\lambda \sum_{i}v_i\sigma_ip_i,
\end{eqnarray}
was used in the derivation of the response function $\chi(\omega,\mathbf{q})$. Here $\lambda=\pm $ denotes the chirality of Weyl nodes,
$\sigma_i$ are the Pauli matrices, $\mathbf{p}=-i\bm{\nabla}$ is the momentum operator, $\hbar$ and the light velocity $c$ are taken to be unit throughout this Letter. 
The description of the axial gauge fields $A_{5,0}$ and $\mathbf{A}_5$, in general, requires more complicated models where there is a separation between Weyl nodes in momentum space and/or energy. For example, $\mathbf{A}_5$ appears in a Weyl semimetal with a broken TRS. We stress, however, that the AMEE is general and does depend on the presence of ${\bf E}_5$ but not the origin of the axial gauge fields: these gauge fields can be generated by light, strain or other means. In addition, we notice that all nodes contribute additively to the magnetization (see also SM~\cite{SI}).

Magnetization given in Eq.~\eqref{M-def} with the response function provided in Eq.~\eqref{Eq:IFE} represents the first main result of this Letter. The response function $\chi(\omega,\mathbf{q})$ can be applied in both clean and dirty limits and is also valid for two-dimensional gapless or gapped Dirac semimetals. 
Furthermore, the result in Eq.~\eqref{Eq:IFE} is applicable to both conventional inverse Faraday and the axial magnetoelectric effects in Dirac materials.

\begin{figure}
 	\includegraphics[width=.4\textwidth]{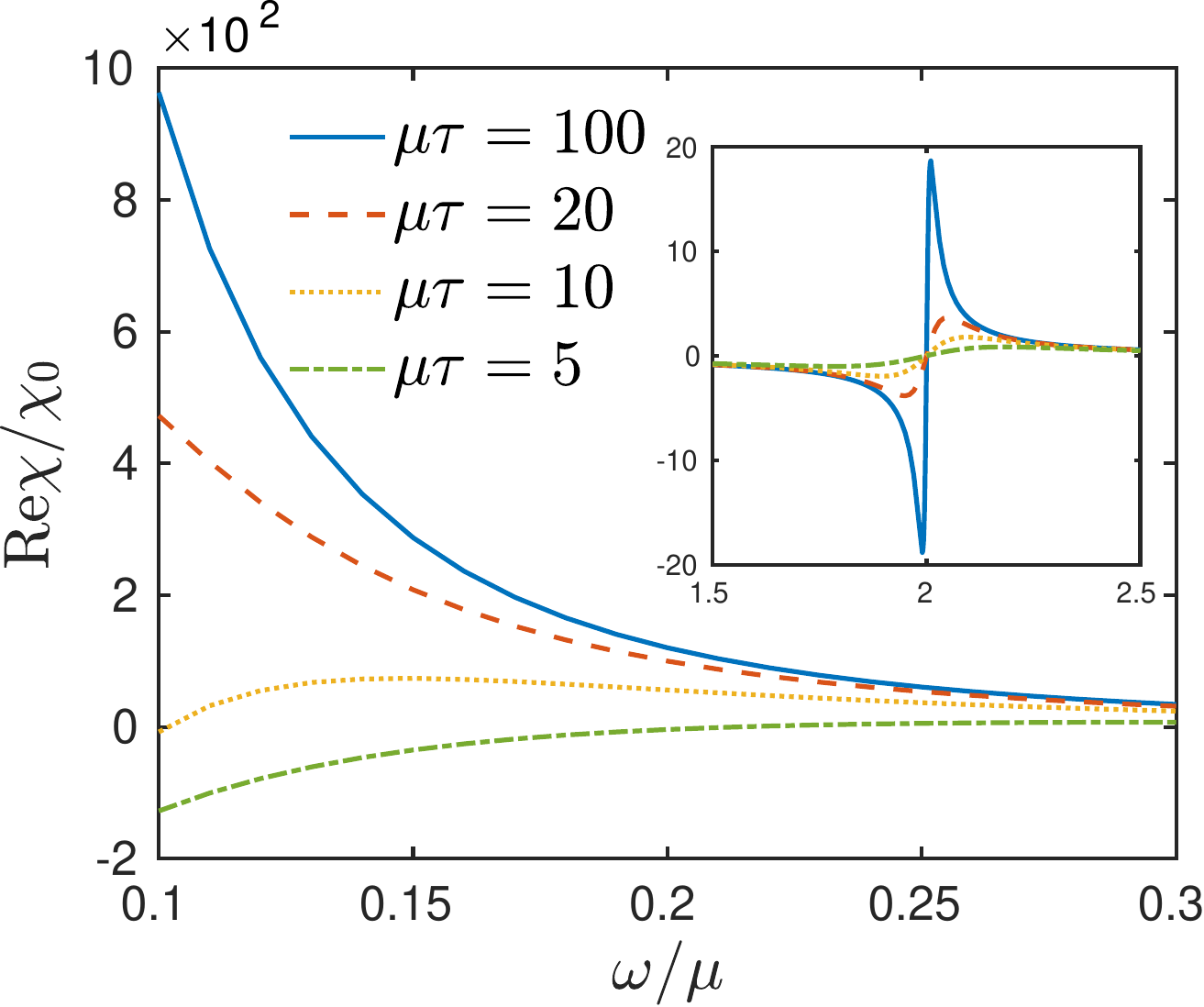}
 	\caption{The frequency dependence of the real part of the normalized response function $\chi/\chi_0$ for different scattering times. The inset shows the results for large frequencies $\omega\sim2\mu$, where the interband effects are important. Here $\chi_0=e^3v_xv_y/(6\pi^2v_z\mu^2)$, $\mu$ is the chemical potential and $\tau$ is the scattering time.}\label{Fig:chi}
 \end{figure}

For a single Weyl node, we find in the $\mathbf{q}=0$ limit
\begin{eqnarray}\label{Eq:chi_Weyl}
\chi=\frac{e^3v_x v_y \mu}{6\pi^2v_z\omega}\left[
\frac{1}{\left(\omega+\frac{i}{\tau}\right)^2}+\frac{3}{\left(\omega+2\mu+\frac{i}{\tau}\right)\left(\omega -2\mu+\frac{i}{\tau}\right)}\right],\nonumber\\
\end{eqnarray}
where $e$ is the electric charge, $\mu$ is the chemical potential, and $\tau$ is the scattering time. Figure~\ref{Fig:chi} shows the real part of $\chi$ as a function of $\omega/\mu$ for different scattering times.
In the $\omega\tau\gg 1$ limit, the first term in Eq.~\eqref{Eq:chi_Weyl}, which is the intraband contribution, reduces to the result obtained using the semiclassical theory~\cite{PhysRevB.101.174429}. In the dirty limit, $\omega\tau\ll 1$, the intraband term changes sign. Note that in this limit the semiclassical treatment does not apply. The interband contribution diverges at $\omega=2\mu$ in the clean limit, and this divergence disappears when the scattering time is finite. In contrast, $\chi$ diverges in the small $\omega$ limit even for finite scattering time.
The induced magnetization vanishes when the chemical potential crosses exactly the nodal point. The reason is that the system has particle-hole symmetry at this point. Under particle-hole transformation, the current changes sign, while the axial current remains unchanged~\cite{Peskin:1995ev}, and therefore the correlation function $\langle j_{5,\mu}j_{5,\nu}j_{\rho} \rangle$ that determines the magnetization (see Fig.~\ref{Fig:FeynmanDiagram} and SM~\cite{SI}) vanishes identically.

\emph{Strain-induced AMEE.--} As an example, we calculate the AMEE in strained Dirac semimetals. The axial gauge field is related to the strain fields as~\cite{PhysRevLett.115.177202}
\begin{equation}
A_{5,i}=\frac{b}{e}\left[\beta u_{iz}+\tilde{\beta}(b)\delta_{iz}\sum_{j}u_{jj}\right],
\end{equation}
where the Weyl nodes are separated along the $z$-direction, $\beta$ and $\tilde{\beta}(b)$ are related to the Gr\"{u}neisen constants, and $u_{ij}=(\partial_i u_j +\partial_j u_i)/2$ is the strain tensor with $\mathbf{u}$ being the displacement field. 
As an example, we consider a circularly polarized sound wave propagating in the $z$-direction,
\begin{equation}
\mathbf{u}=\mathrm{Re} \left[u_0(\mathbf{e}_x-i\mathbf{e}_y) e^{i (q_z z-\omega t)}\right],
\end{equation}
where $\mathbf{e}_x$ and $\mathbf{e}_y$ are unit vectors in the $x$- and $y$-directions as well as $\omega=v_s q_z$ with the transverse sound velocity $v_s$. 

Then the axial vector potential is 
\begin{equation}
\mathbf{A}_5=i(\mathbf{e}_x-i\mathbf{e}_y)\frac{b\beta u_0}{4e}q_ze^{i (q_z z-\omega t)}+\mathrm{c.c.},
\end{equation}
which gives rise to a circularly polarized axial electric field.
Note that since $A_5\propto q_z\propto \omega$,  the axial electric field is proportional to $\omega^2$.

Since the sound velocity is much smaller than the Fermi velocity, $\omega\ll v_F q_z$,
we can no longer take the $q_z=0$ limit as for conventional IFE. 
Because the sound frequency is small compared to the relaxation timescale, the most experimentally relevant regime is $\omega\tau\ll 1$. On the other hand, since $v_z/v_s\gg 1$, the condition $q_z v_z\tau\ll 1$ can be easily violated.
For example, if we take $\omega=1$\si{GHz} and $v_z/v_s=100$, then $q_zv_z\tau=1$ for $\tau=10$\si{ps}.  As a comparison, the transport scattering time can be as long as $0.2$\si{ns} in the Dirac semimetal Cd$_3$As$_2$~\cite{Cd3As2_Ong}. 
Therefore, it is necessary to consider both $q_zv_z\tau\ll 1$ and $q_zv_z\tau\gg 1$ limits.

The intraband contribution to the magnetization in the $\omega\tau\ll 1$ and $q_zv_z\tau\ll 1$ limit is
\begin{equation}
M^{\mathrm{intra}}_z\approx M_{0}\left[1-\frac{3}{5}(q_z v_z\tau)^2\right]\mu \omega \tau^2, \label{Eq:AcousticIFE_intra}%\propto \omega^3.
\end{equation}
and in the $\omega\tau\ll 1$ and $q_zv_z\tau\gg 1$ limit, we derive
\begin{equation}
M^{\mathrm{intra}}_z\approx 3M_{0} \frac{1}{(q_z v_z\tau)^2}\left[1-\frac{\pi}{2(q_z v_z\tau)}\right]\mu \omega\tau^2.\label{Eq:AcousticIFE_intra_large_qz}
\end{equation}
Here $M_{0}=e v_x v_y (\beta b u_0 q_z)^2/(48\pi^2 v_z)\propto (u_0\omega)^2\propto I$, where $I$ is the intensity of the sound (see, e.g., Ref.~\cite{Abrikosov}). In Eq.~\eqref{Eq:AcousticIFE_intra}, we have expanded the result up to $\omega (q_zv_z)^2$ and omitted the subleading term $\propto \omega^3$. To obtain the above results, we assumed that $\mu\tau$ is large enough and $q_zv_z\tau\ll \mu\tau$. See SM~\cite{SI} for details. The interband contribution becomes important only for a small chemical potential or a short relaxation time $\mu\tau\ll 1$.

\begin{figure}
	\includegraphics[width=0.4\textwidth]{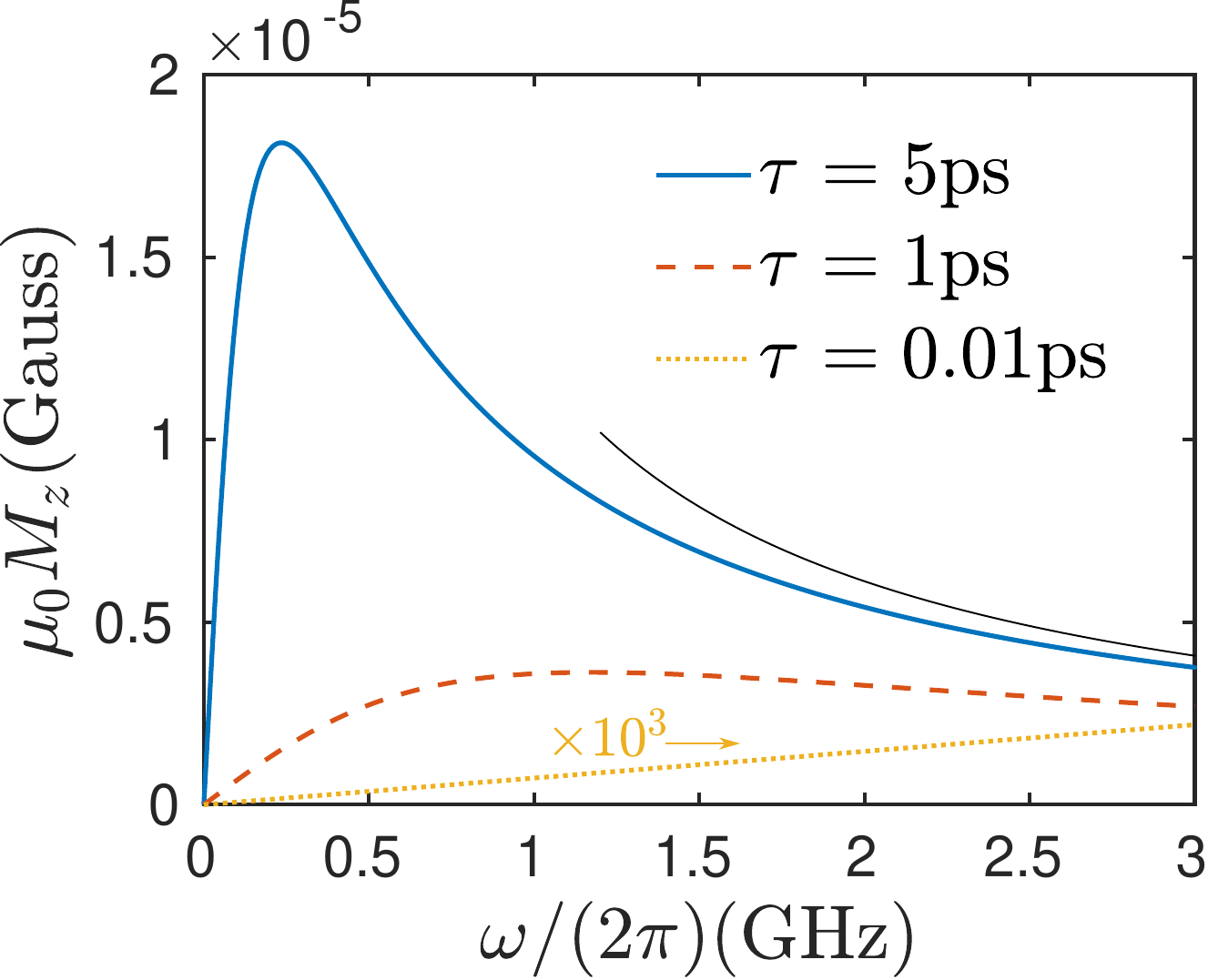}
	\caption{The frequency dependence of the induced magnetization for Cd$_3$As$_2$ for different scattering times at the fixed sound intensity. Here $\mu_0$ is the vacuum permeability. The results for $\tau=0.01$\si{ps} are multiplied by $10^3$. The chemical potential  is taken to be $0.2$\si{eV}, and the sound intensity is $10$\si{W/cm^2}. The black thin curve shows the magnetization calculated using the first term in Eq.~\eqref{Eq:AcousticIFE_intra_large_qz}.
}\label{Fig:Acoustic_IFE}
\end{figure}

At large frequencies ($q_zv_z\tau\gg 1$), magnetization $M^{\mathrm{intra}}_z$ scales as $\mu/\omega$, while for the small ones ($q_zv_z\tau\ll 1$), $M^{\mathrm{intra}}_z\propto\mu\omega\tau^2$. This scaling is different from the case of the IFE~\cite{PhysRevB.101.174429} and is related to a different frequency dependence of the axial gauge fields. It is worth emphasizing that the AMEE is different also in the strong dependence on the wave vector $q_z$, which is negligible for the IFE.

Another notable feature is related to a presence of a peak in the magnetization. The peak appears due to the momentum dependence of the magnetization. 
From Eqs.~\eqref{Eq:AcousticIFE_intra} and \eqref{Eq:AcousticIFE_intra_large_qz}, we can see that the magnetization reaches a maximum at $\omega\tau\propto v_s/v_z$ and the peak value is proportional to $\mu\tau$.
The scaling behavior and the appearance  of a peak of the sound-induced magnetization at small frequencies are in drastic contrast to the IFE, where the magnetization is proportional to $1/\omega^3$ in the clean limit and to $\tau^2/\omega$ in the dirty limit. The scaling of the AMEE with $\omega$ should be robust as long as the strains can be interpreted as effective axial gauge fields and the deviations from the relativisticlike spectrum are small.

To estimate the effect, we use parameters for the Dirac semimetal Cd$_3$As$_2$~\cite{Cd3As2_Chen,2007MAW200717,Cd3As2_SoundVelocity}: $v_x=8.47$\si{eV\AA}, $v_y=8.56$\si{eV\AA}, $v_z=2.16$\si{eV\AA}, $b=0.16$\si{\AA^{-1}}, and $v_s=1.6\times 10^{3}$\si{m/s}, which is about $200$ times smaller than $v_z$.
The chemical potential is fixed to be $0.2$\si{eV}. 
We further assume that $\beta=1$~\footnote{%Gr\"{u}neisen 
Gruneisen 
constants are relatively material-independent and are of order $1-2$~\cite{Ibach-Luth:book}} and the sound intensity is $I=10$\si{W/cm^2} (see, e.g., Refs.~\cite{Weinreich-White:1957,Smith-Sproule:1959}).
In Fig.~\ref{Fig:Acoustic_IFE}, we show the numerical results for the frequency dependence of the magnetization at different scattering times. The black thin line is plotted using the first term in Eq.~\eqref{Eq:AcousticIFE_intra_large_qz}, $M_z=3M_0\mu\omega/(q_zv_z)^2\propto 1/\omega$.

For $\tau=0.01$\si{ps}, which is of the same order as the quantum lifetime measured in Cd$_3$As$_2$~\cite{Cd3As2_Ong}, the magnetization scales linearly with $\omega$ and the slope is proportional to $\tau^2$, and for $\tau=5$\si{ps}, it scales as $1/\omega$ for $\omega/(2\pi)\gtrsim 2$\si{GHz}. The peak is clearly observed for $\tau=5$\si{ps}. 
The positions and heights of the peaks are in agreement with our estimates in Eqs.~\eqref{Eq:AcousticIFE_intra} and \eqref{Eq:AcousticIFE_intra_large_qz}. 
For $\tau=1$\si{ps} at $\omega/(2\pi)=0.2$\si{GHz},  the generated magnetic field strength is $\mu_0 M_z\approx 1$\si{\mu G}. Thus the magnetic flux through a sample of diameter $1$\si{mm} is about one-tenth of the flux quantum, which is five orders above the SQUID threshold~\cite{SQIUD,Granata_2016}. With the decrease of $\tau$, the magnetization decreases significantly, but it is still detectable even for $\tau=0.01$\si{ps} at $\omega/(2\pi)=0.1$\si{GHz}.
The interband correction to the magnetization for $\tau=1$\si{ps} and $\tau=5$\si{ps} is negligible, while for $\tau=0.01$\si{ps}  it contributes about $7\%$ of the total magnetization in the plotted frequency region. Therefore, as one might naively expect, for large chemical potentials, the interband effects play a minor role in the sound-induced processes. 

Unlike the IFE, where the wave vector of light is negligibly small, $q_z$ can be of the same order as the Fermi wave vector $\mu/v_z$ for the AMEE. Thus, electrons can backscatter on the Fermi surface at $q_zv_z\sim 2\mu$ for strong (pseudo)spin-flipping scattering potentials, giving rise to a second peak in the induced magnetization. See SM~\cite{SI} for details.

\emph{Conclusion.--} 
We propose a dynamical axial magnetoelectric effect where a static magnetization is generated as a result of the transfer of the angular momentum of the axial gauge field ${\bf E}_5(t)$ to the orbital magnetization of the electron quasiparticles. The resulting magnetization is of the second order in dynamical fields and reads as $\mathbf{M} \sim \mathbf{E}_5(\omega)\times \mathbf{E}_5^*(\omega)$.

The  proposed axial magnetoelectric  effect  is  universal  with respect to the origin of the dynamic axial fields and could appear in many systems. 
As an example, we propose to use dynamical deformations (sound) in Dirac and Weyl semimetals to excite these fields and generate static magnetization. We found that, unlike the IFE, the induced magnetization scales as $\omega$ ($1/\omega$) rather than $1/\omega$ ($1/\omega^3$) for small (large) frequencies in the dirty limit and at fixed sound intensity. 

By using realistic model parameters, we estimated the induced magnetization in the Dirac semimetal Cd$_3$As$_2$.
Being within experimental reach, the effect provides a way to investigate unusual axial electromagnetic fields via conventional magnetometry techniques.

\begin{acknowledgments}
{\it Acknowledgments.--} We are grateful to M. Geilhufe for useful discussions. This work was  supported  by  Nordita, the  University  of  Connecticut,  and the  European  Research  Council  under  the  European  Unions Seventh Framework ERC-2018-SYG 810451 HERO, VILLUM  FONDEN  via  the  Centre  of  Excellence  for  Dirac Materials (Grant No.  11744) and KAW. POS acknowledges the support through the Yale Prize Postdoctoral Fellowship in Condensed Matter Theory. 
\end{acknowledgments}

%\bibliography{AMEE.bib}
\input{AMEE.bbl}
\end{document}

\title{Supplemental material to ``Axial magnetoelectric effect in Dirac semimetals"
}

\newcommand{\affiliationNordita}{Nordita, KTH Royal Institute of Technology and Stockholm University, Roslagstullsbacken 23, SE-106 91 Stockholm, Sweden}

\newcommand{\affiliationConnecticut}{Department of Physics, University of Connecticut, Storrs, Connecticut 06269, USA}

\newcommand{\affiliationYale}{Department of Physics, Yale University, New Haven, CT 06520, USA}

\author{Long Liang}
\affiliation{\affiliationNordita}
\author{P. O. Sukhachov}
\affiliation{\affiliationYale}
\author{A. V. Balatsky}
\affiliation{\affiliationNordita}
\affiliation{\affiliationConnecticut}

\maketitle

\section{Derivations of the axial magnetoelectric effect}

In this section, we derive the axial magnetoelectric effect (AMEE). To calculate the induced magnetization, we apply a testing magnetic field to the system and integrate out the fermions to get the effective action in terms of the axial gauge fields and testing magnetic field. Then the magnetization $\mathbf{M}$ is obtained as a variation of the effective action with respect to the field, $\mathbf{M}=-\lim_{B\to 0}\delta S/\delta \mathbf{B}$. This approach is valid when the perturbations are weak such that the system is close to equilibrium.
The relevant part of the effective action reads
\begin{eqnarray}
S=\int\mathrm{d}r_1\mathrm{d}r_2\mathrm{d}r_3~\langle j^\mu(r_1)j^\nu_5(r_2) j^\rho_5(r_3) \rangle A_{\mu}(r_1)A_{5,\nu}(r_2)A_{5,\rho}(r_3).
\end{eqnarray}
The diagrammatic expression for the magnetization is shown in Fig.~2 in the main text.
The testing magnetic field $\mathbf{B}$ is generated through a vector potential $\mathbf{A}$. The vector potential couples to the vector current $\mathbf{j}$ via the standard coupling $\propto \mathbf{j}\cdot\mathbf{A}$. The axial potential $\mathbf{A}_5$, which is responsible for the dynamic field $\mathbf{E}_5$, couples to the axial current $\mathbf{j}_5$ as $\propto \mathbf{j}_5\cdot\mathbf{A}_5$. Thus, to calculate the second-order in axial gauge fields contribution to the magnetization, one needs to evaluate the correlation function of two axial and one vector current operators. In addition, as explained in the main text, the contribution to the magnetization from each node simply add up.

It could be instructive to notice that this is not the case for the photogalvanic current, where the contributions from the Weyl nodes of opposite chiralities have opposite signs. This will be evident from the expressions below.

The orbital magnetization for a single node is
\begin{eqnarray}
M_z=e^3\mathrm{Tr}\int \mathrm{d}r\mathrm{d}k\mathrm{d}p\mathrm{d}q~
\left[r_x\hat{v}_y-r_y \hat{v}_x\right]g_{k}\mathbf{A}_{5,q}\cdot \hat{\mathbf{v}}g_{k-q}
\mathbf{A}_{5,p}\cdot\hat{\mathbf{v}}g_{k-p-q}e^{i(p+q)r},
\end{eqnarray}
where $\hat{\mathbf{v}}=\nabla_{\mathbf{k}}H(\mathbf{k})$ is the velocity operator. The factor $1/(2\pi)^d$ is absorbed in the measure. For monochromatic axial gauge field $\mathbf{A}_{5}$ with frequency $\omega$ and momentum $\mathbf{q}$, the integration over $p$ and $q$ becomes the sum over discrete frequency and momentum $\pm (i\omega, \mathbf{q})$. The calculations are performed by using the imaginary-time formalism, with $r=(\uptau,\mathbf{r})$,   $k=(i\omega_n,\mathbf{k})$, and the Matsubara Green's function given by
\begin{eqnarray}
g(i\omega_n,\mathbf{k})=\sum_m\frac{|u_{m,\mathbf{k}}\rangle\langle u_{m,\mathbf{k}} |}{i\omega_n-\xi_{m,\mathbf{k}}+\frac{i}{2\tau}\mathrm{sgn}(\omega_n)}.
\end{eqnarray}
Here $m=\pm$ denotes the particle and hole band, $|u_{m,\mathbf{k}}\rangle$ is the eigenstate of the Hamiltonian with $H(\mathbf{k})|u_{m,\mathbf{k}}\rangle=\epsilon_{m,\mathbf{k}}|u_{m,\mathbf{k}}\rangle$, $\xi_m=\epsilon_{m}-\mu$ with $\mu$ being the chemical potential, and $\tau$ is a phenomenological relaxation time that describes the dissipation. Note that the velocity operator for Dirac or Weyl semimetals is momentum independent. If the Hamiltonian contains terms quadratic in momentum, then there will be extra diagrams in the magnetization.

We focus on the static magnetization, which means $q+p=0$.  Without loss of generality, we assume that the axial electric field is in the $x-y$ plane. Then the magnetization reads
\begin{eqnarray}
M_z
&=&e^3 \sum_{l,m,n}\int \mathrm{d}\mathbf{r}\mathrm{d}\mathbf{k}~ \langle u_{n,\mathbf{k-p-q}}|r_x\hat{v}_y-r_y\hat{v}_x|u_{m,\mathbf{k}}\rangle \langle u_{m,\mathbf{k}}|\hat{v}_i |u_{l,\mathbf{k-q}}\rangle\langle u_{l,\mathbf{k-q}}|\hat{v}_j|u_{n,\mathbf{k-p-q}}\rangle e^{i\mathbf{(p+q)\cdot r}}A_{5,i}(i\omega,\mathbf{q})A_{5,j}(-i\omega,\mathbf{p})\nonumber\\
&&~~
\sum_{\omega_r}\frac{1}{i\omega_r-\xi_{m,\mathbf{k}}+\frac{i}{2\tau}\mathrm{sgn}(\omega_r)}\frac{1}{i\omega_r-i\omega-\xi_{l,\mathbf{k-q}}+\frac{i}{2\tau}\mathrm{sgn}(\omega_r-\omega)}\frac{1}{i\omega_r-\xi_{n,\mathbf{k-p-q}}+\frac{i}{2\tau}\mathrm{sgn}(\omega_r)} +\left(\begin{array}{c}
\omega\to-\omega \\
\mathbf{q}\to -\mathbf{q} \\
\mathbf{p}\to -\mathbf{p}
\end{array}\right).
\nonumber\\
\end{eqnarray}
If we directly take $\mathbf{p+q}=0$, then the integral simply becomes zero because the integrand in this limit is  an odd function of the position $r_x$ and $r_y$. To avoid this problem, we replace $r_x$ by $\sin(Qr_x)/Q$, and $r_y$ by $\sin(Qr_y)/Q$, and take the $Q\to 0$ limit after integrating over $\mathbf{r}$. Physically, this means that we use a spatially varying testing magnetic field and take the uniform limit after the calculation. This approach agrees with that in Ref.~\cite{Shi-Niu:2007}.

Since there are only two bands, the possible combinations of $m,l,n$ are $m=n=l$, $m=n\ne l$, $m=l\ne n$, and $n=l\ne m$. The $m=n=l$ terms are intraband processes. For $m=n\ne l$, the state is excited to a different state and then goes back to the initial state, for the other two cases the initial state is different from the final state.

We first consider the clean limit with $\mathbf{q}=0$, then the summation over Matsubara frequencies can be carried out directly. We find that for $l=m=n$, 
\begin{eqnarray}
\label{M-intra-def}
M^\mathrm{intra}_z=-ie^3\frac{[\mathbf{A}_5(\omega)\times\mathbf{A}_5(-\omega)]_z}{2}\sum_m \int\mathrm{d}\mathbf{k}~\frac{f'_m}{\omega}
\left[
v^m_y v^m_x\partial_{k_x}v^m_y-v^m_yv^m_y \partial_{k_x}v^m_x+(x\leftrightarrow y)
\right],
\end{eqnarray}
where $v^m_i=\partial_{k_i}\epsilon_m$, $f_m=1/(e^{\xi_m/T}+1)$ is the Fermi-Dirac distribution function and $f'_m=\partial_{\epsilon_m} f_m$. 
In the above equation we have switched to the real frequency. Since this term comes from intraband processes, it is dubbed the intraband magnetization. For $m=n\ne l$, we find
\begin{eqnarray}
\label{M-inter-def}
M^\mathrm{inter}_z=-ie^3\frac{[\mathbf{A}_5(\omega)\times \mathbf{A}_5(-\omega)]_z}{2}\sum_{m\ne l} \int\mathrm{d}\mathbf{k}\frac{\omega f'_m (\epsilon_l-\epsilon_m)}{\omega^2-(\epsilon_l-\epsilon_m)^2}
\left[(\epsilon_l-\epsilon_m)^2(\Omega^m_{xy})^2+(v^m_xv^m_xg^m_{yy}-v^m_xv^m_yg^m_{xy})+(x\leftrightarrow y) \right],\nonumber\\
\end{eqnarray}
where $\Omega^m_{xy}=i\langle \partial_{k_x} u_{m,\mathbf{k}}|\partial_{k_y} u_{m,\mathbf{k}}\rangle -(x\leftrightarrow y)$  and $g^m_{ij}=\sum_{l\ne m}\langle \partial_{k_i} u_{m,\mathbf{k}} |u_{l,\mathbf{k}}\rangle\langle u_{l,\mathbf{k}} | \partial _{k_j} u_{m,\mathbf{k}}\rangle+(i\leftrightarrow j)$
are the Berry curvature~\cite{Berry:1984} and the quantum metric~\cite{Provost1980} of the band $m$, respectively. For a generic two-band model with the Hamiltonian $H=\mathbf{h}\cdot\bm{\sigma}$, we have~\cite{PhysRevB.95.024515}
\begin{eqnarray}
\label{Omega-def}
\Omega^{\pm}_{ij}&=&\pm\frac{1}{2}\hat{\mathbf{h}}\cdot (\partial_{k_i} \hat{\mathbf{h}} \times \partial_{k_j}\hat{\mathbf{h}}),\\
\label{g-def}
g^{\pm}_{ij}&=&\frac{1}{2}\partial_{k_i} \hat{\mathbf{h}}\cdot\partial_{k_j} \hat{\mathbf{h}},
\end{eqnarray}
with $\hat{\mathbf{h}}=\mathbf{h}/|\mathbf{h}|$ being a unit vector. 
The term given in Eq.~(\ref{M-inter-def}) involves interband processes. Therefore, we call it the interband magnetization. The other processes sum to zero. Note that both the intra- and interband contribution contains the derivative of the Fermi-Dirac distribution, so the AMEE is a Fermi surface property. This also explains why the separation between Weyl nodes should not enter the corresponding magnetizations. Indeed, such anomalous terms usually originate from the states deeply below the Fermi surface (see e.g., Ref.~\cite{Gorbar:2017wpi}). 
Finally, we ignored the terms in the Hamiltonian that break the particle-hole symmetry $\propto h_0$. In the vicinity of the Dirac nodes, these terms shift chemical potential and lead to the dependence of the velocity on the band index. Both effects provide only a quantitative correction to the intraband contribution, which is dominant in the AMEE for the parameters used in the main text.

For nonzero scattering time, we can write the Matsubara frequency summation as a contour integral~\cite{altland_simons_2010}. We find 
\begin{eqnarray}
\label{M-general-def}
M_z=-i\chi(\omega,\mathbf{q})[\mathbf{E}_5(\omega,\mathbf{q})\times\mathbf{E}_5(-\omega,-\mathbf{q})]_z,
\end{eqnarray}
with the response function
\begin{eqnarray}
\chi(\omega,\mathbf{q})&=&\frac{e^3}{4\omega^2}\Bigg\{\sum_m \int\mathrm{d}\mathbf{k}~I_{mmm}(\omega,\mathbf{q})
\left(
v^m_y v^m_x\partial_{k_x}v^m_y-v^m_yv^m_y \partial_{k_x}v^m_x
\right)\nonumber\\
&&+\sum_{m\ne l} \int\mathrm{d}\mathbf{k}~I_{mml}(\omega,\mathbf{q})(\epsilon_l-\epsilon_m)\left[(\epsilon_l-\epsilon_m)^2(\Omega^m_{xy})^2+ v^m_xv^m_xg^m_{yy}-v^m_xv^m_yg^m_{xy} \right]+(x\leftrightarrow y)
\Bigg\}.\label{Eq:AME_chi}
\end{eqnarray}
Here
\begin{eqnarray}
&&I_{mnl}(\omega,\mathbf{q})\nonumber\\
&=&\int\frac{\mathrm{d}x}{2\pi i}\left\{\frac{f(x+\omega)-f(x)}{(x-\xi_{m,\mathbf{k}}-\frac{i}{2\tau})(x+\omega-\xi_{l,\mathbf{k+q}}+\frac{i}{2\tau})(x-\xi_{n,\mathbf{k}}-\frac{i}{2\tau})}\right.
-\frac{f(x)-f(x-\omega)}{(x-\xi_{m,\mathbf{k}}+\frac{i}{2\tau})(x-\omega-\xi_{l,\mathbf{k-q}}-\frac{i}{2\tau})(x-\xi_{n,\mathbf{k}}+\frac{i}{2\tau})}\nonumber\\
&&+\frac{f(x)}{(x-\xi_{m,\mathbf{k}}+\frac{i}{2\tau_m})(x+\omega-\xi_{l,\mathbf{k-q}}+\frac{i}{2\tau_l})(x-\xi_{m,\mathbf{k}}+\frac{i}{2\tau_n})}
-\frac{f(x-\omega)}{(x-\xi_{m,\mathbf{k}}+\frac{i}{2\tau})(x-\omega-\xi_{l,\mathbf{k-q}}+\frac{i}{2\tau})(x-\xi_{n,\mathbf{k}}+\frac{i}{2\tau})}\nonumber\\
&&+\frac{f(x)}{(x-\xi_{m,\mathbf{k}}-\frac{i}{2\tau})(x-\omega-\xi_{l,\mathbf{k-q}}-\frac{i}{2\tau})(x-\xi_{n,\mathbf{k}}-\frac{i}{2\tau})}
-\left.\frac{f(x+\omega)}{(x-\xi_{m,\mathbf{k}}-\frac{i}{2\tau})(x+\omega-\xi_{l,\mathbf{k+q}}-\frac{i}{2\tau})(x-\xi_{n,\mathbf{k}}-\frac{i}{2\tau})}\right\}.
\end{eqnarray}
To get the above result we have assumed that $\mathbf{q}$ is small such that the wave function $|u_{\mathbf{k-q}}\rangle$ can be expanded as $|u_{\mathbf{k-q}}\rangle\approx |u_{\mathbf{k}}\rangle - |\bm{\nabla}_{\mathbf{k}}u_{\mathbf{k}}\rangle \cdot\mathbf{q}$.
At vanishing temperature, the integration over $x$ can be calculated analytically.
Introducing 
\begin{eqnarray}
J_1(a,b)&=&\int^0_{-\infty}\mathrm{d}x~\frac{1}{(x-a-\frac{i}{2\tau})^2(x-b+\frac{i}{2\tau})},\\
J_2(a,b)&=&\int^0_{-\infty}\mathrm{d}x~\frac{1}{(x-a+\frac{i}{2\tau})^2(x-b+\frac{i}{2\tau})},
\end{eqnarray}
then $\chi(\omega,\mathbf{q})$ can be written in terms of $J_1$ and $J_2$ with 
\begin{eqnarray}
J_1(a,b)&=&\frac{2}{(2a+\frac{i}{\tau})(a-b+\frac{i}{\tau})}+\frac{\ln{\left(\frac{1+4b^2\tau^2}{1+4a^2\tau^2}\right)}-i2\pi+2i \left[\arctan{(2a\tau)}+\arctan{(2b\tau)}\right]}{2(a-b+\frac{i}{\tau})^2},\\
J_2(a,b)&=&\frac{2}{(a-b)(2a-\frac{i}{\tau})}+\frac{\ln{\left(\frac{1+4b^2\tau^2}{1+4a^2\tau^2}\right)}+2i \left[\arctan{(2b\tau)}-\arctan{(2a\tau)}\right]}{2(a-b)^2}.
\end{eqnarray}
These expressions are used in numerical calculations later.

%Results for Weyl and 2D Dirac
\section{The AMEE for Weyl and Dirac fermions}

In this section, we study the AMEE for Weyl and two-dimensional Dirac fermions.

\subsection{AMEE for Weyl fermions}
The dispersion for the Weyl fermion is $\epsilon_{\pm}=\pm\sqrt{v_i^2k_i^2}\equiv \pm \epsilon$. The Berry curvature for the Weyl node with chirality $\lambda=\pm$ is
\begin{eqnarray}
\Omega^{m}_{xy}= \lambda \frac{v_x v_y  v_z k_z}{2\epsilon^3_m},
\end{eqnarray}
and the quantum metric is
\begin{eqnarray}
g^m_{ij}=\frac{\epsilon^2\delta_{ij} - v_iv_j k_i k_j}{2\epsilon^4}v_iv_j.
\end{eqnarray}
These expressions follow from Eqs.~\eqref{Omega-def} and \eqref{g-def}.
It is worth noting that both intraband and interband contributions to the magnetization given in Eqs.~(\ref{M-intra-def}) and (\ref{M-inter-def}) are chirality-even functions. Indeed, in the model at hand, only the Berry curvature explicitly depends on chirality. However, while the intraband contribution does not depend on the Berry curvature, the interband one is even in $\Omega^{m}_{xy}$. Therefore, the contributions to the magnetization from each node simply add up.

For large $\mu\tau$, it is difficult to numerically calculate the integral in Eq.~\eqref{Eq:AME_chi} with high accuracy because the integrand is singular-like (the most important region is around the Fermi surface). But it is possible to find analytical expressions. 
To proceed we scale $v_ik_i$ to $k_i$ and $v_iq_i$ to $q_i$, then in terms of the new variables $\epsilon_\pm=\pm k$ and 
\begin{eqnarray}
\chi(\omega,\mathbf{q})&=&-\frac{e^3v_xv_y}{2v_z\omega^2}\Bigg[ \int\mathrm{d}\mathbf{k}~(I_{+++}-I_{---})
\frac{k_x^2+k_y^2}{k^3}+ \int\mathrm{d}\mathbf{k}~(I_{++-}-I_{--+})\frac{k^2+3k_z^2}{k^3}
\Bigg].\label{Eq:IFE_disspition_Weyl}
\end{eqnarray}
We further rotate the wave vector $(v_x q_x, v_yq_y, v_z q_z)$ to be along the $z$-direction, i.e., $(v_x q_x, v_yq_y, v_z q_z)\equiv (0,0,q)$ and expand $\epsilon_{\mathbf{k-q}}$ as $\epsilon_{\mathbf{k-q}}=k-q \cos{\theta}$, where $\theta$ is the angle between $\mathbf{k}$ and the $z$-axis. For the above expansion to be valid, we assumed that $\mu\tau$ is large such that the most important contribution is from the Fermi surface and $q\tau\ll \mu\tau$.  
By introducing the shorthand notations  $I_{+++}(\omega,q)=I_1(k,\cos{\theta},\omega,q)$, $I_{++-}(\omega,q)=I_2(k,\cos{\theta},\omega,q)$,  
$I_{---}(\omega,q)=I_1(-k,-\cos{\theta},\omega,q)$, $I_{--+}(\omega,q)=I_2(-k,-\cos{\theta},\omega,q)$, and $y=\cos{\theta}$, we rewrite Eq.~(\ref{Eq:IFE_disspition_Weyl}) as 
\begin{eqnarray}
\chi(\omega,\mathbf{q})&=&-\frac{e^3v_xv_y}{8\pi^2 v_z\omega^2} \int^\infty_0k\mathrm{d}k\int^{1}_{-1} \mathrm{d}y ~\Bigg\{\left[I_{1}(k,y,\omega,q)-I_{1}(-k,-y,\omega,q)\right]
(1-y^2)\nonumber\\
&&+ \left[I_{2}(k,y,\omega,q)-I_{2}(-k,-y,\omega,q)\right](1+3y^2)
\Bigg\}.\label{Eq:IFE_disspition_Weyl2}
\end{eqnarray}
Because the above integral is symmetric with respect to $y$, the integrand is an even function of $k$. Therefore, Eq.~(\ref{Eq:IFE_disspition_Weyl2}) can be rewritten as
\begin{eqnarray}
\chi(\omega,\mathbf{q})&=&-\frac{e^3v_xv_y}{8\pi^2 v_z\omega^2} \int^\infty_{-\infty}k\mathrm{d}k\int^{1}_{-1} \mathrm{d}y ~\Bigg[I_{1}(k,y,\omega,q)
(1-y^2)+ I_{2}(k,y,\omega,q)(1+3y^2)
\Bigg].\label{Eq:IFE_disspition_Weyl3}
\end{eqnarray}
Then the integration over $k$ can be done with the help of the residue theorem. At vanishing temperature, the integration over $x$ in $I_1$ and $I_2$ can also be performed analytically. We find
\begin{eqnarray}
\chi(\omega,\mathbf{q})&=&\frac{e^3v_xv_y}{16\pi^2 v_z\omega^2} \int^{1}_{-1} \mathrm{d}y ~\Bigg[\frac{2\mu\omega(1-y^2)}{(\omega-qy+i/\tau)^2}
-
\frac{(2\mu+q y)\omega(1+3y^2)}{(2\mu+q y +\omega+i/\tau)(2\mu+q y -\omega-i/\tau)}
\Bigg].\label{Eq:AME_Weyl_large_mu}
\end{eqnarray}
In the $q=0$ limit, we obtain
\begin{eqnarray}\label{Eq:chi_Weyl}
\chi(\omega,\mathbf{0})=\frac{e^3v_x v_y \mu}{6\pi^2v_z\omega}\left[
\frac{1}{(\omega+\frac{i}{\tau})^2}+\frac{3}{(\omega+2\mu+\frac{i}{\tau})(\omega -2\mu+\frac{i}{\tau})}\right],
\end{eqnarray}
which is given in Eq.~(4) in the main text. The first term in the square bracket is the intraband contribution and the second term the interband one.

\subsubsection{Strain-induced AMEE in Weyl semimetals}

To investigate the stain-induced AMEE, we focus on the $\omega\tau\ll 1$ limit. Indeed, while the typical relaxation time in Dirac and Weyl semimetals is $\tau\sim 0.01-1$\si{ps}, the highest ultrasound frequencies are on a GHz scale.
By expanding Eq.~\eqref{Eq:AME_Weyl_large_mu} up to the first order of $\omega\tau$, we find the following intraband term:
\begin{eqnarray}
\chi^\mathrm{intra}\approx-\frac{e^3v_xv_y}{2\pi^2 v_z\omega^2}\frac{q\tau-\arctan{q\tau}}{(q\tau)^3}\mu\omega\tau^2,
\end{eqnarray}
which in the $q\tau \ll 1$ limit is,
\begin{eqnarray}
\chi^\mathrm{intra}\approx-\frac{e^3v_x v_y }{6\pi^2v_z\omega^2}\bigg[1-\frac{3}{5} (q\tau)^2\bigg]\mu\omega\tau^2
\end{eqnarray}
and for $q\tau \gg 1$ reads as
\begin{eqnarray}
\chi^{\mathrm{intra}}\approx-\frac{e^3v_x v_y }{2\pi^2v_z\omega^2} \frac{1}{(q\tau)^2}\bigg[1-\frac{\pi}{2q\tau}\bigg]\mu\omega\tau^2.
\end{eqnarray}
Note that here $q$ is the rescaled wave vector. In the main text, we use $q=q_z v_z$.

The interband response function in the $\omega\tau\ll 1$ limit reads,
\begin{eqnarray}
\chi^\mathrm{inter}&=&-\frac{e^3v_x v_y }{16\pi^2v_z\omega^2}
\frac{\omega\tau}{q^3 \tau^3}\bigg\{12 \mu \tau \big[
\arctan{(q\tau - 2 \mu\tau) } + 
\arctan{(q\tau + 2 \mu\tau)}-q \tau \big] \\
&&+ \big[ (q^2 + 
12 \mu^2) \tau^2-3\big] \arctanh{\left(\frac{4 q \mu \tau^2}{
1 + (q^2 + 4 \mu^2) \tau^2}\right)}\bigg\},
\end{eqnarray}
which in the $q\tau \ll 1$ limit is
\begin{eqnarray}
\chi^\mathrm{inter}=-\frac{e^3v_x v_y }{2\pi^2v_z\omega^2}\bigg[\frac{1}{1+4\mu^2\tau^2} + \frac{7(4\mu^2\tau^2-3)q^2\tau^2}{15 (1+4\mu^2\tau^2)^3} \bigg]\mu\omega\tau^2.
\end{eqnarray}

The above analytical results apply for large $\mu\tau$. For small $\mu\tau$, $q\tau$ can easily exceed $\mu\tau$. In this case, the contribution to the integral is not concentrated around the Fermi surface. Therefore, we can not simply expand $\epsilon_{\mathbf{k-q}}$ as $k-q\cos{\theta}$. However, the integrand in Eq.~\eqref{Eq:AME_chi} is less singular in this limit  and it is easy to calculate it numerically. 

\begin{figure}
	\includegraphics[width=0.8\textwidth]{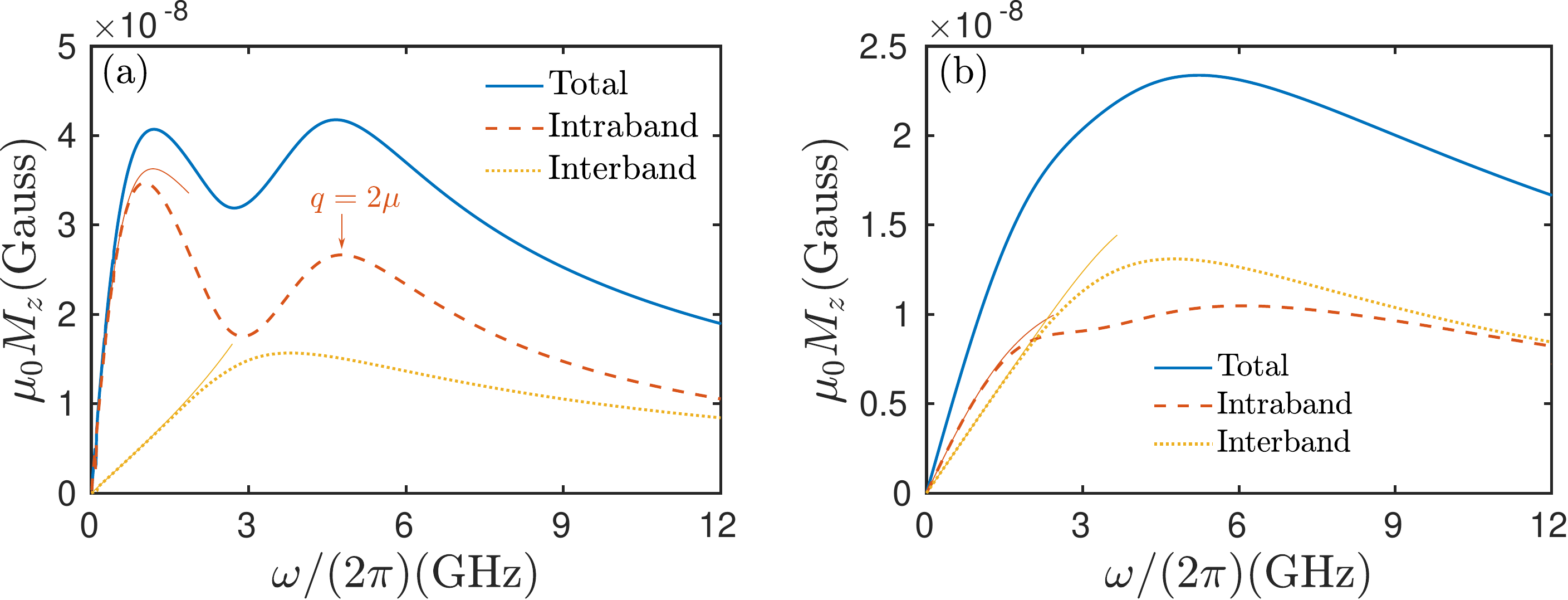}
	\caption{Frequency dependence of the AMEE in Cd$_3$As$_2$ for $\tau=1$\si{ps} (a) and $\tau=0.3$\si{ps} (b). The sound intensity is $10$\si{W/cm^2}. The chemical potential is  $\mu=2$\si{meV}. The second peak in the intraband magnetization is clearly seen in (a), with its location being $q=2\mu$. Here $q=v_z/v_s\omega$ with $v_s$ being the sound velocity and $v_z$ being the Fermi velocity in the $z$-direction. By using the parameters for Cd$_3$As$_2$,  we find $\omega/(2\pi)\approx 4.8$\si{GHz} for $\mu=2$\si{meV}.
	The thin curves are plotted according to Eq.~\eqref{Eq:AME_Weyl_large_mu}.  They agree very well with the full numerical results for small $\omega$.}\label{Fig:AME_small_mu}
\end{figure}

We show the induced magnetization in strained Cd$_3$As$_2$ for small chemical potential $\mu=2$\si{meV} in Fig.~\ref{Fig:AME_small_mu}.  
As in the main text, 
we consider a circularly polarized sound wave propagating in the $z$-direction,
\begin{eqnarray}
\mathbf{u}=\mathrm{Re} \left[u_0(\mathbf{e}_x-i\mathbf{e}_y) e^{i (q_z z-\omega t)}\right],
\end{eqnarray}
where $\mathbf{e}_x$ and $\mathbf{e}_y$ are unit vectors in the $x$- and $y$-directions as well as $\omega=v_s q_z$ with the transverse sound velocity $v_s=1.6\times 10^{3}$\si{m/s}. The Fermi velocities for Cd$_3$As$_2$~\cite{Cd3As2_Chen} are $v_x=8.47$\si{eV\AA}, $v_y=8.56$\si{eV\AA}, and $v_z=2.16$\si{eV\AA}.
The axial vector potential is 
\begin{eqnarray}
\mathbf{A}_5=i(\mathbf{e}_x-i\mathbf{e}_y)\frac{b\beta u_0}{4e}q_ze^{i (q_z z-\omega t)}+\mathrm{c.c.},
\end{eqnarray}
which gives rise to a circularly polarized axial electric field $\mathbf{E}_5=-\partial_t \mathbf{A}_5$. Here the Weyl nodes are separated along the $z$-direction, $\beta$ and $\tilde{\beta}(b)$ are related to the Gr\"{u}neisen constants, and $u_{ij}=(\partial_i u_j +\partial_j u_i)/2$ is the strain tensor with $\mathbf{u}$ being the displacement field. The separation parameter is estimated as~\cite{Cd3As2_Chen} $b=0.16$\si{\AA^{-1}} and we assume that $\beta=\tilde{\beta}(b)=1$.

\begin{figure}
	\includegraphics[width=0.6\textwidth]{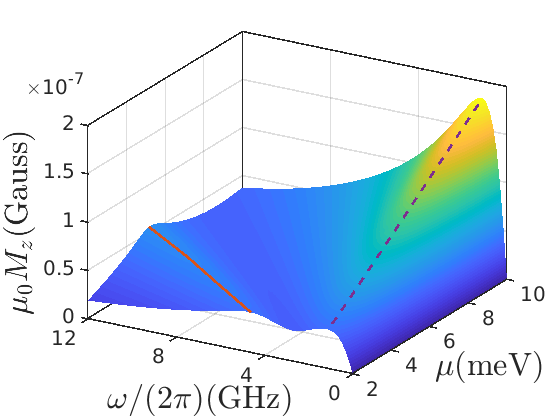}
	\caption{The AMEE magnetization in Cd$_3$As$_2$ as a function of frequency $\omega$ and chemical potential $\mu$. The experimental setup is the same as in the main text.  The scattering time is $\tau=1$\si{ps}. The position of the first peak (the dashed line) is determined by $q\propto 1/\tau$ (see the main text).  The position of the second peak (the solid red line), which appears at higher frequencies, is determined by $q=2\mu$. Here $q=v_z/v_s\omega$ with $v_s$ being the sound velocity and $v_z$ being the Fermi velocity in the $z$-direction. By using the parameters for Cd$_3$As$_2$, we find the position of the second peak is given by $\omega$(\si{GHz})$\approx2\pi\times 2.4 \mu$(\si{meV}). This agrees well with the  numerical results, which give $\omega$(\si{GHz})$\approx 2\pi\times 2.3 \mu$(\si{meV}).
	 }\label{Fig:AME_3Dplot_fixed_tau}
\end{figure}

As shown in Fig.~\ref{Fig:AME_small_mu}~(a), the intraband contribution shows a double-peak feature, which is also visible in the total magnetization. 
The second peak appears because electrons can be backscattered on the Fermi surface at $q\sim 2\mu/v_F$ for appropriate strong scattering potentials. This is also confirmed by the linear dependence of the position of the second peak on the chemical potential shown in Fig.~\eqref{Fig:AME_3Dplot_fixed_tau}.
The peaks are gradually washed out for smaller relaxation times, see Fig.~\ref{Fig:AME_small_mu}~(b). The thin curves in the figure are calculated according to Eq.~\eqref{Eq:AME_Weyl_large_mu}.  As one can see, the result agrees perfectly with the full numerical calculations for small enough $\omega$. 
Notice also that the interband contribution in the AMEE can be comparable in magnitude to that of the intraband one at high frequencies $\omega\sim \mu v_s/v_F$. On the other hand, the interband processes do not lead to any quantitatively different features.

\subsubsection{Crystal symmetries}

It is instructive to briefly discuss the role of crystal point group symmetries for the AMEE. Indeed, the symmetries restrict the form of the response tensor $\chi_{jl}$ in a generalized analog of Eq.~(\ref{M-general-def}),
\begin{equation}
M_j=-i \chi_{jl}(\omega, \mathbf{q}) \left(\mathbf{E}_{5,\omega,\mathbf{q}}\times \mathbf{E}_{5,-\omega,-\mathbf{q}}\right)_l.
\end{equation}
Here, we already used the fact that both $\mathbf{M}$ and $\left(\mathbf{E}_{5,\omega,\mathbf{q}}\times \mathbf{E}_{5,-\omega,-\mathbf{q}}\right)$ are pseudovectors.

Depending on the crystal structure, the Dirac semimetal Cd$_3$As$_2$ has a $C_{4v}$ or $D_{4h}$ point group symmetry~\cite{Wang:2013}.
Diagonal components of the response tensor $\chi_{jl}$ do not break mirror, parity-inversion, and time-reversal symmetries. Furthermore, $\sum_{j}\chi_{jj}$ is a scalar for the point group symmetries. In the main text, we focused on the diagonal component $\chi_{zz}$, which is not restricted by $C_{4v}$ and $D_{4h}$ symmetries. 
As for off-diagonal components of $\chi_{jl}$, they require breaking two mirror symmetries with respect to the mirror planes normal to the magnetization and the sound wave vector. This requirement is inconsistent with the presence of two mirror planes in the $C_{4v}$ point group symmetry. The other point group symmetry, $D_{4h}$, also does not allow for the off-diagonal elements. Thus, only diagonal components of the response tensor $\chi_{jl}$ survive in Cd$_3$As$_2$, which does not contradict our analysis in a low-energy effective model.

% Dirac
\subsection{AMEE for 2D Dirac fermions}

Let us consider the case of two-dimensional (2D) Dirac fermions. The formula for the AMEE magnetization is similar to the 3D Weyl semimetal case (see Eqs.~\eqref{M-general-def} and \eqref{Eq:IFE_disspition_Weyl}), with $k_z$ being replaced by the mass or gap $M$. We find in the clean limit,
\begin{eqnarray}
M_z=i\theta(|\mu|-|M|)\mathrm{sgn}(\mu)\frac{v_x v_y e^3}{4\pi\omega^2}\left[\frac{\mu^2-M^2}{\mu^2\omega}
+\frac{\omega}{\omega^2-(2\mu)^2}\frac{\mu^2+3M^2}{\mu^2}\right] \left[\mathbf{E}_{5,\omega}\times \mathbf{E}_{5,-\omega}\right]_z,
\end{eqnarray}
where $\theta(x)$ is the step function. In the massless limit, this result reduces to the magnetization derived in Ref.~\cite{PhysRevB.101.174429}. It is worth noting also that, the interband contribution comes solely from the quantum metric in the massless case since the Berry curvature vanishes. In contrast, both the Berry curvature and quantum metric contributions are nonzero for Weyl semimetals. 

In the dirty limit, we find
\begin{eqnarray}
M_z^\mathrm{intra}&=&-i\frac{v_x v_y e^3}{\pi\omega^2}\left[\frac{M^2\mu \tau^2}{\omega^2}
+\frac{8}{3\pi}\mu\omega\tau^3\right] \left[\mathbf{E}_{5,\omega}\times \mathbf{E}_{5,-\omega}\right]_z,\\
M_z^\mathrm{inter}&=&-i\frac{v_x v_y e^3}{\pi\omega^2}\frac{8}{3\pi}\mu\omega\tau^3 \left[\mathbf{E}_{5,\omega}\times \mathbf{E}_{5,-\omega}\right]_z,
\end{eqnarray}
where the first and second equations are the intraband and interband contributions, respectively.

%\bibliography{AMEE.bib}
\input{supplement.bbl}